# Investigating the accuracy of the diffusion coefficient estimation based on the rotating ring-disk electrode transient measurements in the cases of kinetic-limited regime and disk surface passivation.

Artem V. Sergeev[1*], Alexander V. Chertovich[2,1]

[1] Physics Department, Lomonosov Moscow State University, Moscow, 119991 Russia

[2] Semenov Institute of Chemical Physics, Moscow, 119991 Russia

* Corresponding author. Physics Department, Lomonosov Moscow State University, Leninskie Gory, Moscow, 119991 Russia. Tel.: +7 495 939 4013. E-mail address: sergeev@polly.phys.msu.ru

**Abstract**

The chronoamperometric measurements performed with rotating ring-disc electrode become a one of the standard techniques for the diffusion coefficient determination. However, the theoretical estimation of the transit time on which the technique is based is not strictly connected to the apparent time interval measured in the experiments. Therefore, we employed numerical simulation of RRDE in order to predict possible error of the diffusion coefficient estimation in respect to the disc electrode conditions. According to the results the obtained estimation can vary up to two times depending on the disk electrode potential. On the other hand, even an intense disk passivation process leading to the peak-like shape of the ring current transient does not affect significantly the apparent transit time and the diffusion coefficient estimation. Additionally, it was showed, that the presence of a linear segment in the disk current transient indicates that the time it took to cover most of the disk electrode with the passivation film is close to the RRDE characteristic time.

**Introduction**

A rotating ring-disk electrode (RRDE) originally was invented by Frumkin and Nekrasov[1] almost half a century ago. The addition of the ring electrode made it possible to detect the reaction product generated at the disk as it moved outward the electrode axis by the convection flow with controlled speed. Such design proved to be effective to analyze even short-living species. The RRDE become a tools widely used for a wide range of research problems such as electrochemical water splitting[2–4], oxidation of organic molecules[5,6], plating[7,8] and stripping[9,10] of metals and alloys, reduction of oxygen[11–14] and sulfur[15]. Although voltamperometric experiments might be more common, chronoamperometric data obtained with RRDE were often used, particularly, to estimate the diffusion coefficient[16,17] of the disc reaction product.

The technique of diffusion coefficient determination with the help of RRDE is based on the theoretical estimate of the transit time[18] (that is needed for the disk reaction product to reach the ring) $T_{trans} = (\lg r_2/r_1)^{2/3}(\nu/D)^{1/3}/\omega$, where $D$ is diffusion coefficient, $\nu$ kinematic viscosity, $\omega$ is electrode rotation speed, $r_1$ and $r_2$ are disc and internal ring radii respectively. In practice the transit time is determined by extrapolation of the ring current transient linear region[19]. However, the shape and slope of the ring current transient depends on the disk current amplitude in the first moments of the experiment. Thus in case of kinetic or mixed regime the apparent transit time determined in such a way may depend on the disk electrode potential. The error produced by the uncertainty of the disk electrode initial condition is yet to be established.

The original derivation of the transit time expression also does not imply any surface changes in time (e.g. passivation process). Although the disk passivation process can alter the ring transient shape significantly up to the point at which it degrade into a peak[20] that should affect the apparent transit time. That is an actual problem for the Li-air battery research. At the Li-air cathode two electrochemical reactions proceed simultaneously: oxygen reduction reaction $(O_2 + e^- + Li^+ \rightarrow LiO_2)$[21–24] and secondary lithium superoxide reduction to lithium peroxide $(LiO_2 + e^- + Li^+ \rightarrow Li_2O_2 \downarrow)$[25–27]. The first reduction produces the flux of soluble $LiO_2$ that can be detected by electrooxidation at the ring. The latter leads to passivating $Li_2O_2$ film growth.

In this work we employed numerical simulation to investigate passible variation of the diffusion coefficient estimation based on the disk-to-ring transit time measured with the help of RRDE in respect to disk electrode potential (i.e. diffusion-limited, mixed and kinetic regimes). Furthermore, we performed analogous study in the case of disk electrode passivation. The results obtained provide error estimation and help to choose the experiment conditions in order to increase the accuracy of the diffusion coefficient measurements done with RRDE.

**Computational details**

Originally the RRDE numerical models were developed by Prater and Bard[28]. The information on the recent progress in this area can be found e.g. in [29]. In this work a 2-D mathematical model of RRDE system was developed and applied. The model implements a numerical solution of differential equations describing diffusion and convection phenomena in the framework of the finite differences method. The system of convection-diffusion equations[30] relies on the approximate analytical solution for the flow velocities derived by Karman[31] and Cochran[32]. The spatial grid step along the radial coordinate (in the plane of the electrode surface) was chosen to be 25 µm. In all the simulations we used 100 grid points along the normal axis (perpendicular to the surface), which covered distance equal to 1.6 of the diffusion layer thickness $\delta_0$ ($\delta_0 = 1.61 D^{1/3} v^{1/6} \omega^{-1/2}$). Thus the normal direction grid step depended on the diffusion coefficient of the disk reaction product, the solution kinematic viscosity and the electrode rotation speed. The time step was also related the characteristic time of the system $\tau = 0.51^{-2/3}(v/D)^{1/3}\omega^{-1}$, so there were 1000 steps per $\tau$.

The basic version of the model was modified in order to simulate kinetic and mixed regimes according to equation $1/I_{disk} = 1/I_{kin} + 1/I_{diff}$, where $I_{kin}$ is the kinetic current under certain disk potential and surface reagent concentration equal to the bulk concentration, $I_{diff}$ is diffusion-limited current according to Levich equation. In order to investigate the effect of passivation on the apparent values of transit time the island growth mechanism was implemented in the RRDE model. The passivation model implies covering the surface by the insolating film of certain thickness (i.e. finite capacity $Q_{max}$ in terms of charge consumed). Therefore, the film growth diminishes the fraction of active surface: $d\frac{A}{A_{max}} = -\frac{\beta I_{disk}}{Q_{max}}dt$, where β is the fraction of disc current consumed by passivation process, $A_{max}$ – initial total active disk electrode surface.

A simple redox system i.e. ferrocyanide/ferricyanide was used as a reference case for choosing the simulation parameters. Particularly 5 mM $[Fe(CN)_6]^{3+}/[Fe(CN)_6]^{4+}$ in 1M aqueous solution of KCl. The ferrocyanide(ferricyanide) diffusion coefficient was set to $D$ = 0.67(0.72) cm$^2$/s, kinematic viscosity $v$ = 0.009 cm$^2$/s). The electrode dimensions were: disk $d_1$= 4 mm, ring inner diameter $d_2$= 5 mm, ring outer diameter $d_2$= 7 mm. The error of the calculated collection efficiency turned out to be about 1% in comparison to the theoretical expression[33].

**Results and Discussion**

The simulation study was started from the investigation of how the disk electrode regime (diffusion-limited, mixed or kinetic) affects the measured transit time and diffusion coefficient estimation. That was done by varying the ratio of kinetic and diffusion currents $I_{kin}/I_{diff}$ that represents the variation of the disc electrode potential. The simulated disk currents are presented in Figure 1. The pure diffusion-limited regime takes place at $I_{kin}/I_{diff}$ equal to positive infinity (+Inf), though it can be seen that in practice any value $I_{kin}/I_{diff} > 10$ yield the same disk current curve. The characteristic feature of the diffusion-limited regime is a current spike at the start of the experiment (simulation) that decays to steady state value within 0.2 s according (which we refer as the RRDE characteristic time). That is in a good agreement with other theoretical predictions and numerous experimental data. At $I_{kin}/I_{diff} = 1/3$ the disc current is almost constant that manifests kinetic regime. The regime that set in at values $1/3 < I_{kin}/I_{diff} < 10$ should be considered as mixed. It should be noted, that the steady state current value becomes smaller for the lower kinetic current values. However, RRDE is a linear system, thus scaling of the disc current can not affect the shape of the ring current, but only results in the same linear scaling.

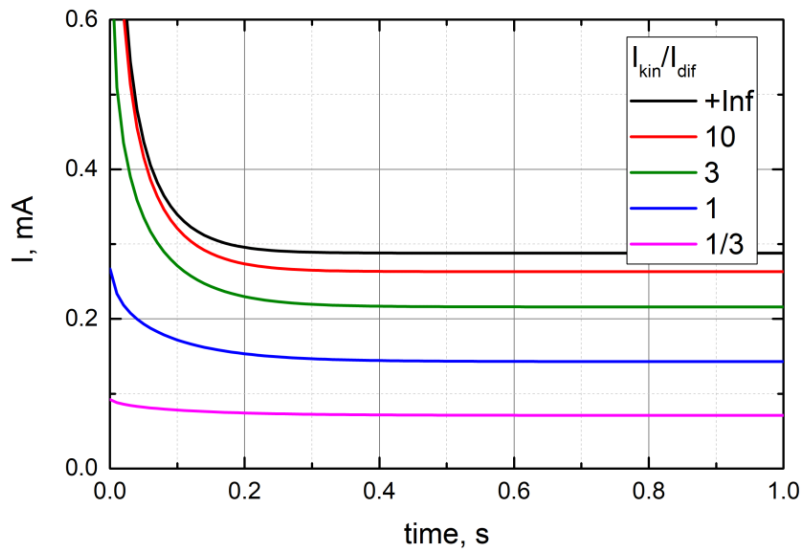

*Figure 1. Disk currents simulated for the different values of the kinetic to diffusion current ration in the range from infinity (diffusion-limited regime). The electrode rotation speed is 800 rpm.*

The ring currents obtained in the same series of simulation are presented in Figure 2 along with the lines fitted to the linear regions of the ring current transients. The transients obtained at the higher kinetic current values are steeper due to emission of extra amount of product at the disk just after switching the potential on. The apparent transit time turned out to be substantially different as well as the calculated diffusion coefficient values which are given in Table 1. The diffusion coefficient value estimated under diffusion-limited condition is 1.3*10$^{-5}$ cm$^2$/s, that almost 2 times higher than the true value 6.7*10$^{-6}$ cm$^2$/s used in the simulation (the error upon increasing the fitting range to 60% of full amplitude is no more than +5%). On the other hand, the kinetic regime simulation ($I_{kin}/I_{diff} = 1/3$) yielded the diffusion coefficient estimation of 7.4*10$^{-6}$ cm$^2$/s which is in a good agreement with the true value. According to this results the kinetic regime at the disk (i.e. low overpotential) seems to be more preferable.

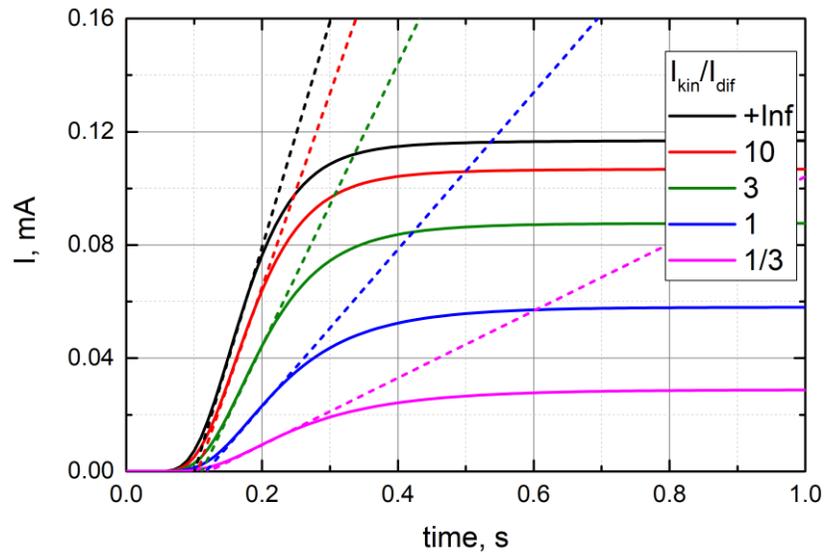

*Figure 2. Ring currents simulated for the different values of the kinetic to diffusion current ration in the range from infinity (diffusion-limited regime). Dashed lines are the extrapolations of the transient linear regions used to determine the transit times. The electrode rotation speed is 800 rpm.*

In order to investigate the effect of passivation on the apparent values of transit time the island growth mechanism was implemented in the RRDE model. The simulations were carried out for a set of passivation current fraction values β in the range from 0 to 10%. The full capacity of the passivating film $Q_{max}$ was set to be equal to the diffusion limiting current multiplied by 0.02s (which is about the RRDE characteristic time needed to reach the steady state regime at the disk. The initial $I_{kin}/I_{diff}$ ratio was set to 3 (mixed regime at the disk), though the effective value of this parameter has been decreasing in the course of the simulation due to surface passivating that manifested in disk current fading (as seen in Figure 3). Passivation current fraction β > 5% resulted in full passivation in time comparable to the RRDE characteristic time and peak-shaped ring currents (see Figure 4). Furthermore, stronger passivation leads to smaller peak current values. Nevertheless, the apparent transit times turned out almost the same. The diffusion coefficient calculated basing on transit time in case of severe passivation (β = 10%) is only 10% higher than the value obtained without passivation effect (β = 0). All the calculated diffusion coefficient values can be found in Table 1.

*Table 1. Diffusion coefficient values calculated based on the transit times obtained by varying different simulation parameters.*

| Parameter | Value | | | | |
|---|---|---|---|---|---|
| $I_{kin}/I_{dif}$ | +Inf | 10 | 3 | 1 | 1/3 |
| D, cm$^2$/s*10$^{-4}$ | 1.30 | 1.14 | 0.96 | 0.82 | 0.74 |
| passivation fraction β | 0 | 0.5% | 2% | 5% | 10% |
| D, cm$^2$/s*10$^{-4}$ | 0.96 | 0.96 | 0.97 | 1.00 | 1.05 |
| rotation speed, rpm | 200 | 400 | 800 | 1200 | 1800 |
| D, cm$^2$/s*10$^{-4}$ | 1.17 | 1.07 | 1.00 | 0.96 | 0.93 |

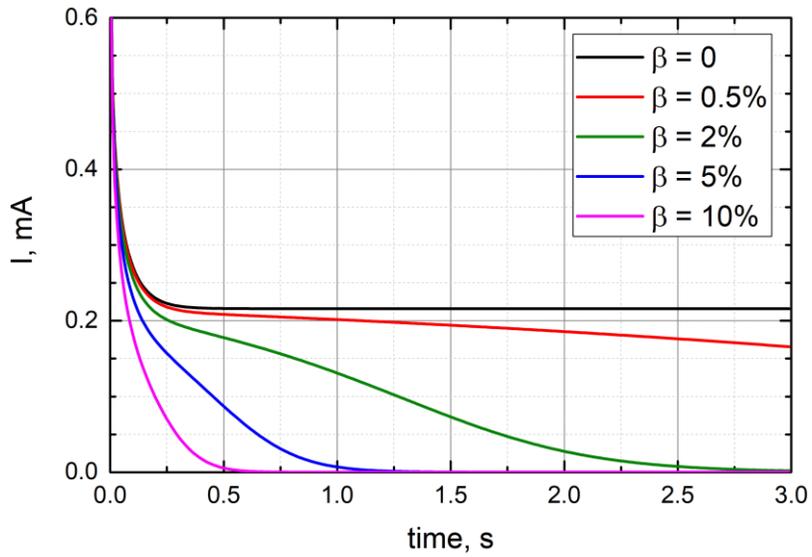

*Figure 3. Disk currents simulated for different values of the passivation current fraction. The electrode rotation speed is 800 rpm.*

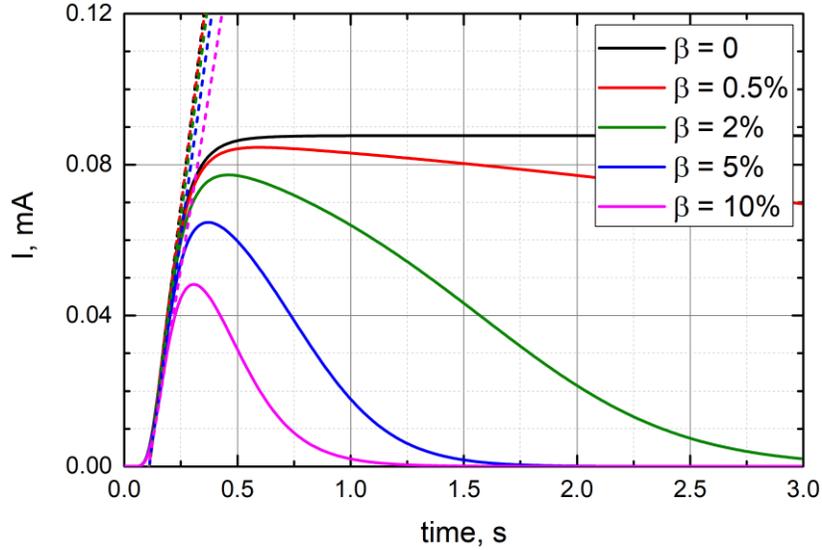

*Figure 4. Ring currents simulated for different values of the passivation current fraction. Dashed lines are the extrapolations of the transient linear regions used to determine the transit times. The electrode rotation speed is 800 rpm.*

Under high overpotential (high kinetic current) the disk current is limited by diffusion, so that partial surface passivation should not affect it until most of the surface is covered by insulating film. Therefore, we investigated cases of different passivating film capacity values, as it determines the time needed to fill the disk surface. The simulations were carried out for passivating film $Q_{max}$ values equal to the diffusion limiting current ($I_{dif}$ = 2.8 mA) multiplied by 5, 20 and 80 ms. The passivation current fraction β set to 5%; the initial $I_{kin}/I_{diff}$ ratio was set to 3. The simulated disk currents are presented in Figure 5, while the evolution of disk surface active (unpassivated) is presented in Figure 6.

In case of $Q_{max}$ = $I_{dif}$*5 ms disk current fades slowly for the first 2 seconds, while the active area decreases linearly to the 20 %. After that the kinetic regime sets in and imposes linear dependency of the disk current on the active area. Therefore, both disk current and surface area fades according to exponential law $\frac{A}{A_{max}} \sim \exp\left(-\frac{\beta I_{kin}}{Q_{max}} dt\right)$, where $I_{kin}$ is the kinetic current value under clean surface condition. Nevertheless, the time needed to cover most of the surface with passivating byproduct is an order of time larger that the RRDE characteristic time. Thus the two processes (setting of the diffusion quasi-steady state and surface passivation) can be easily decoupled.

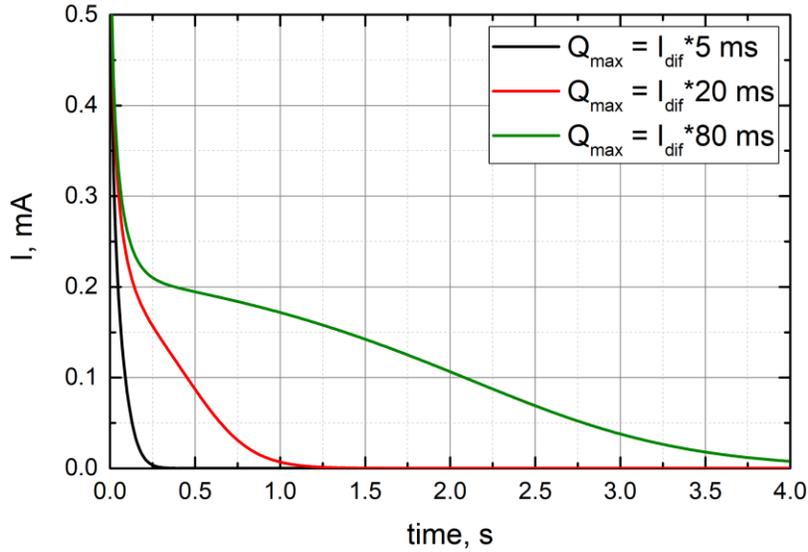

*Figure 5. Disk currents simulated for different values of the passivation film capacity. The electrode rotation speed is 800 rpm.*

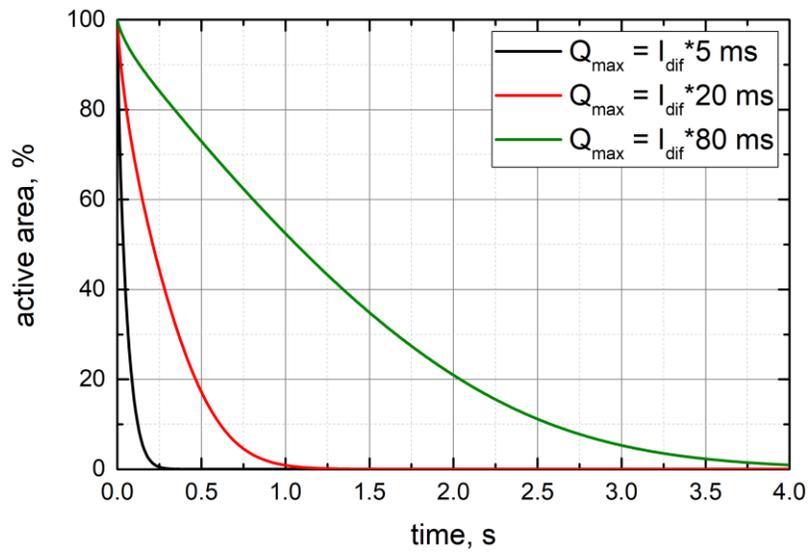

*Figure 6. Electrode active surface area simulated for different values of the passivation film capacity. The electrode rotation speed is 800 rpm.*

In case of $Q_{max}$ = $I_{dif}$*20 ms 80% of surface area deactivates within 0.5 seconds which is comparable to the RRDE characteristic time. That results in a specific shape of the disc current, particularly a prominent linear region. Finally, at $Q_{max}$ = $I_{dif}$*5 ms the complete passivation is achieved faster that the disk current could reach the diffusion-limited value. The disk current linear region disappears in this case. Thus the disk current shape alone can indicate the speed of the passivation process.

Finally, we checked the effect of the electrode rotation speed variation on the diffusion coefficient estimation in the presence of surface passivation. The simulations were done for a range of rotation speeds from 200 to 1800 rpm. The passivation current fraction β set to 5%; the passivating film capacity $Q_{max}$ was set to $I_{dif}$*20 ms. The initial kinetic current was fixed to be $I_{kin}$ ratio was set to 3 times higher that the diffusion limit at 800 rpm rotation speed.. The resulted ring currents are presented in Figure 7 along with the lines extrapolating the transients linear regions used to determine the transit time.

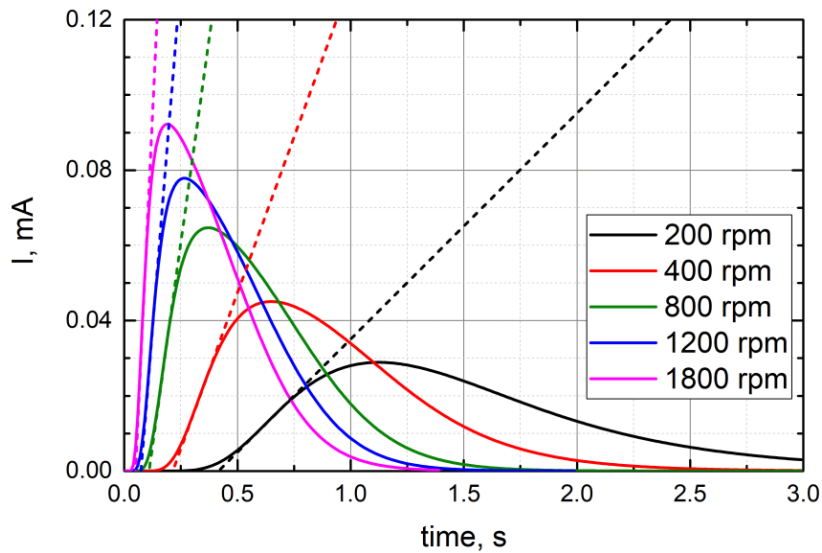

*Figure 7. Ring currents simulated for different values of electrode rotation speed. Dashed lines are the extrapolations of the transient linear regions used to determine the transit times.*

The variation of the calculated diffusion coefficient in respect to electrode rotation speed was only about 10% (see Table 1). The estimation done for higher rotation speed is close to the true value, because for the fixed kinetic current (i.e. fixed electrode potential) the $I_{kin}/I_{diff}$ is smaller, thus the system is closer to the kinetic regime that established to be preferable. Though the diffusion coefficient estimation deference is not significant in the whole range of the rotation speeds.

**Conclusions**

With the help of the numerical model of RRDE it was established that the diffusion coefficient estimation based on the apparent transit time obtained by extrapolation of the ring current transient linear region may be overestimated by up to two times if high overpotential is applied to the disk electrode (diffusion-limited regime is imposed). On the other hand, in case of kinetic-limited regime at the disk the estimation is only 10% higher that the true value. Basing on the simulation of the disk surface passivation phenomenon it was concluded, that even severe passivation that leads to a peak-like ring current shape does not affect significantly the diffusion coefficient estimation based on the transit time. Finally, it was found out, that the presence of a linear segment in the disk current transient indicates that the passivation of most of the desk electrode surface takes time comparable to the RRDE characteristic time needed to reach the steady-state disk current in the absence of passivation.

**Acknowledgements**

The work was financially supported by the Centre for Electrochemical Energy of Skolkovo Institute of Science and Technology.